\documentclass[iop,aplj]{emulateapj}
\usepackage{apjfonts}
\slugcomment{To appear in ApJ Letters; Received 2013 May 8; accepted 2013 May 13.}
\newcommand{\swift}{{\em Swift}}
\newcommand{\chandra}{{\em Chandra}}
\newcommand{\fermi}{{\em Fermi}}
\newcommand{\nustar}{{\em NuSTAR}}
\newcommand{\sgr}{SGR\,J1745$-$29}
\newcommand{\sgrastar}{Sgr\,A*}

\shorttitle{Swift Discovery  of \sgr.}
\shortauthors{Kennea et al.}

\begin{document}

\title{\textit{Swift} discovery of a new Soft Gamma Repeater, \sgr, near Sagittarius A*}

\author{J.~A.\ Kennea\altaffilmark{1}, D. N. Burrows\altaffilmark{1}, C.\
  Kouveliotou\altaffilmark{2}, D.~M.\ Palmer\altaffilmark{3}, E.\
  G\"o\v{g}\"u\c{s}\altaffilmark{4}, Y.\ Kaneko\altaffilmark{4}, P.~A.\
  Evans\altaffilmark{5}, N.\ Degenaar\altaffilmark{6}, 
  M.~T.\ Reynolds\altaffilmark{6}, J.~M.\ Miller\altaffilmark{6}, R.\
  Wijnands\altaffilmark{7}, K.~Mori\altaffilmark{8} and N.\ Gehrels\altaffilmark{9}} 

\affil{$^1$ Department of Astronomy \& Astrophysics, The Pennsylvania State
  University, 525 Davey Lab, University Park, PA 16802, USA; \textit{kennea@swift.psu.edu}}

\affil{$^2$ Science \& Technology Office, ZP12, NASA/Marshall Space Flight Center,
  Huntsville, AL 35812, USA}

\affil{$^3$ Los Alamos National Laboratory, Los Alamos, NM 87545, USA}

\affil{$^4$ Sabanc\i~University, Orhanl\i $-$ Tuzla, \.{I}stanbul 34956, Turkey}

\affil{$^5$ Department of Physics and Astronomy, University of Leicester,
  Leicester, LE1 7RH, UK}

\affil{$^6$ Department of Astronomy, University of Michigan, 500 Church Street,
  Ann Arbor, MI 48109, USA}

\affil{$^7$ Astronomical Institute ''Anton Pannekoek,'' University of Amsterdam,
  Postbus 94249, 1090 GE Amsterdam, The Netherlands}

\affil{$^8$ Columbia Astrophysics Laboratory, Columbia University, New York, NY
  10027, USA}

\affil{$^9$ Astrophysics Science Division, NASA Goddard Space Flight Center, Greenbelt, MD, USA}

\begin{abstract}
  Starting in 2013 February, \swift\ has been performing short daily
  monitoring observations of the G2 gas cloud near \sgrastar\ with the
  X-Ray Telescope to determine whether the cloud interaction leads to an
  increase in the flux from the Galactic center. On 2013 April 24 \swift\
  detected an order of magnitude rise in the X-ray flux from the region
  near \sgrastar. Initially thought to be a flare from \sgrastar, detection
  of a short hard X-ray burst from the same region by the Burst Alert
  Telescope suggested that the flare was from an unresolved new Soft Gamma
  Repeater, \sgr.  Here we present the discovery of \sgr\ by \swift,
  including analysis of data before, during, and after the burst.  We find
  that the spectrum in the 0.3--10\,keV range is well fit by an absorbed
  blackbody model with $kT_\mathrm{BB} \simeq 1$\,keV and absorption
  consistent with previously measured values from the quiescent emission
  from \sgrastar, strongly suggesting that this source is at a similar
  distance. Only one SGR burst has been detected so far from the new
  source, and the persistent light curve shows little evidence of decay in
  approximately 2 weeks of monitoring after outburst. We discuss this light
  curve trend and compare it with those of other well covered SGR
  outbursts. 
  We suggest that \sgr\ belongs to an emerging subclass of magnetars
  characterized by low burst rates and prolonged steady X-ray emission 1-2
  weeks after outburst onset.

\end{abstract}


\section{Introduction}

\citet{Gillessen12} recently reported that a gas cloud referred to as
``G2'' is expected to pass within 3100\,$R_G$ of Sagittarius\,(Sgr)\,A* as
early as mid-2013 \citep{Gillessen13}.  If G2 is indeed a gas cloud
\citep[however, see][]{Phifer13}, its tidal disruption may result in
accretion onto \sgrastar, which in turn could lead to \sgrastar\ entering
an X-ray active state.  The anticipation of this event has led to
monitoring programs of \sgrastar\ over a broad range of wavelengths
starting early 2013.

NASA's \swift\ satellite has a unique rapid slewing capability, which
allows its moderate sensitivity X-Ray Telescope (XRT; \citealt{Burrows05})
to perform short (approximately 1\,ks) daily monitoring of \sgrastar.
Daily observations are being carried out between 2013 February 2 and 2013
November 2, except for a monthly 2--3 day drop out when \sgrastar\ is too
close to the Moon.  Using XRT data from an observation at 17:34 UT on 2013
April 24, \cite{DegenaarATEL5006} reported an increase in the X-ray flux
from the vicinity of \sgrastar\ by an order of magnitude above its
quiescent level. An XRT observation on the previous day did not show any
evidence of enhanced emission from this region. A follow-up observation on
2013 April 25 at 15:58 UT \citep{ReynoldsATEL5016} showed that the enhanced
emission persisted much longer than typical \sgrastar\ flare events, which
only last tens of minutes to hours (e.g., \citealt{Baganoff01,Nowak12}),
making this an unusual flaring episode.

At 19:15 UT on 2013 April 25, during a scheduled observation of \sgrastar,
the \swift\ Burst Alert Telescope (BAT; \citealt{Barthelmy05}) triggered on
a short ($\sim30$\,ms), hard X-ray burst at a position consistent with
\sgrastar\ \citep{BarthelmyGCN14443}.  \cite{KenneaATEL5009} reported that
the characteristics of this burst were consistent with Soft Gamma Repeater
(SGR) bursts seen by BAT, and therefore suggested that both burst and
enhanced emission were from a new SGR source too close to \sgrastar\ for
the XRT (18\arcsec\ HPD, 7\arcsec\ FWHM) to resolve.

SGRs are members of a very small group of sources (26 known to
date\footnote{http://www.physics.mcgill.ca/$\sim$pulsar/magnetar/main.html}),
which are suggested to be magnetars (slowly rotating neutron stars with
extreme surface dipole magnetic fields of $>10^{14}$\,G);
\cite{Duncan92,Kouveliotou98}. Historically, SGRs have been discovered when
they entered a burst active period emitting multiple hard X-ray/soft
$\gamma-$ray bursts at irregular intervals; the first such source was
discovered in 1987 (for reviews on magnetars see
\citealt{Woods06,Mereghetti08} and references therein). All but two magnetars
lie on the Galactic plane with approximately half of their population
concentrated between $\sim7^\circ$ and $\sim30^\circ$ from the Galactic
center.

A \nustar\ follow-up observation on 2013 April 26 found a
$\sim3.76$\,s periodicity \citep{MoriATEL5020}, well within the range of
magnetar periods \citep[2--12\,s;][]{Woods06,Mereghetti08}, further
confirming this source as a likely new SGR.  A subsequent \chandra\
observation on 2013 April 29 found a new X-ray source  $\sim3''$ away
from \sgrastar\ \citep{ReaATEL5032} and confirmed the presence
of the 3.76\,s period. Later \swift\ observations in Windowed Timing mode
allowed a measurement of $\dot{P} = 2.5 \pm 1.1 \times 10^{-11}$,
implying a dipole magnetic field of $B = 3\times10^{14}$\,G, consistent with this
source being a magnetar \citep{GotthelfATEL5046}.
The source was designated \sgr\ \citep{Gehrels13}. 

\sgr\ was observed with the Effelsberg, Green Bank, Parkes and Sardinia
radio telescopes \citep{EatoughATEL5040, BurgayATEL5035, ButtuATEL5053},
which also confirmed the 3.76s period, making \sgr\ the fourth magnetar
detected in radio wavelengths.  \cite{EatoughATEL5040} find a dispersion
measure consistent with \sgrastar .

In this letter we discuss the discovery of \sgr\ by \swift, reporting on the
pre- and post-burst X-ray emission from the source, including spectral and
temporal analyses, and a detailed report on the BAT detection of the SGR
burst. Finally we discuss similarities between \sgr\ and the overall
magnetar population, focusing on the bursting behavior and flux evolution,
and the implications of finding an SGR at the center of our galaxy.  This
letter is a companion to \cite{Mori13}, which describes pulsar timing and broad
band spectral analysis of \sgr\ utilizing primarily \nustar\ data.

\section{Observations}

\swift\ observed the region around \sgrastar\ on an approximately daily
basis beginning 2013 February 03. The detection of increased emission from
the region initiated additional observations through the \swift\ Target of
Opportunity program. As of 2013 May 5, a total of 70.6\,ks of time has been
devoted to observing the \sgrastar\ region with XRT. Further, the
BAT trigger on the \sgr\ burst, resulted in $\sim20$~ks of automated
follow-up observations. Note that observations were not performed on 2013
April 28 to 2013 April 30 due to the field being too close to the Moon for
\swift\ to observe. A summary of the \swift\ observations used in this
letter is given in Table~\ref{observations}.

\begin{deluxetable}{lccr}
\tabletypesize{\scriptsize}
\tablecaption{Log of XRT PC Mode Observations Used in This Letter.\label{observations}}
\tablehead{
\colhead{ObsID}& \colhead{Start Time (UT)}& \colhead{End Time (UT)} & \colhead{Exp.(s)}
}
\startdata
00091736015\tablenotemark{a}& 2013 Apr 20 22:30:02& 2013 Apr 20 22:48:59&    965\\
00091736017\tablenotemark{a}& 2013 Apr 22 17:42:02& 2013 Apr 22 18:01:59&   1030\\
00091736018\tablenotemark{a}& 2013 Apr 23 14:45:02& 2013 Apr 23 16:28:59&    920\\
00091712004	 &2013 Apr 24 17:32:02	 &2013 Apr 24 17:51:58& 1065 \\			
00035650239\tablenotemark{b}	 &2013 Apr 25 14:35:02	 &2013 Apr 25 16:07:57& 995  \\
00091736019	  &2013 Apr 25 19:11:02	 &2013 Apr 25 19:30:58& 1010 \\	   	
00554491000\tablenotemark{c}	  &2013 Apr 25 19:31:02	 &2013 Apr 25 20:49:24& 310 \\	   	
00554491001\tablenotemark{c}	  &2013 Apr 25 20:51:33	 &2013 Apr 26 14:55:54& 19970\\
00091736020	  &2013 Apr 26 05:16:02	 &2013 Apr 26 06:52:59& 1060\\	   	
00035650242\tablenotemark{b}	 &2013 Apr 26 16:03:02	 &2013 Apr 27 16:23:58& 4715\\	   	
00091736021	  &2013 Apr 27 14:28:42	 &2013 Apr 27 14:47:58& 990\\	   	
00091712005	 &2013 May 01 08:13:02	 &2013 May 01 08:31:58& 945\\	   	
00091736022	  &2013 May 02 22:52:02	 &2013 May 02 23:06:11& 710\\	   	
00091736024	&2013 May 04 21:19:02	 &2013 May 04 21:37:59& 970\\	   	
00091736025	  &2013 May 05 23:05:02	 &2013 May 05 23:27:57& 1250\\	   	
00091736026     & 2013 May 06 21:26:02& 2013 May 06 21:48:59 & 1185\\
00091736027     & 2013 May 07 16:41:02& 2013 May 07 16:59:57 & 960\\
00091736028     &2013 May 08 00:30:02& 2013 May 08 00:49:01& 955\\
\enddata
\tablenotetext{a}{Pre-outburst, used for upper limit only.}
\tablenotetext{b}{A target-of-opportunity request.}
\tablenotetext{c}{Observation taken as a result of the BAT trigger on  \sgr.}
\end{deluxetable}

\vskip 160pt
\section{Data Analysis}

We analyzed the \swift\ data with the standard \swift\ analysis tools
version 4.0, part of HEAsoft 6.13.  XRT spectral fitting was performed in
XSPEC \citep{Arnaud96} with the v13 CALDB XRT Photon Counting (PC) mode
RMFs and ARFs.  The ARF files used time-dependent exposure maps to correct
for the presence of hot columns and hot pixels on the total exposure.  All
errors are quoted at 90\% confidence, and coordinates are given in the
J2000 epoch.

\subsection{Rise from Quiescence}

We extracted a light-curve of the region that includes \sgr\ and \sgrastar,
using an extraction region of radius $10''$ centered on the position of
\sgrastar. Compared to the previous quiescent count rates seen from this
region\footnote{http://www.swift-sgra.com}, the data taken between 2013
February 2 and 2013 April 23 show no evidence of enhanced emission, with an
XRT count rate consistent with a non-background subtracted mean of 0.011
s$^{-1}$.

Starting with the observation taken on 2013 April 24 at 17:32 UT,
approximately 1.1 days before the BAT-detected burst, the XRT count rate
from this region had risen to $0.11 \pm 0.1$ s$^{-1}$.  The previous
observation ending 2013 April 16:28 UT showed no evidence of enhanced
emission, and, therefore, we conclude that \sgr\ became active within a
period of $\sim25$ hr.

\subsection{Localization of \sgr}

\begin{figure}
\resizebox{\hsize}{!}{\includegraphics[angle=0]{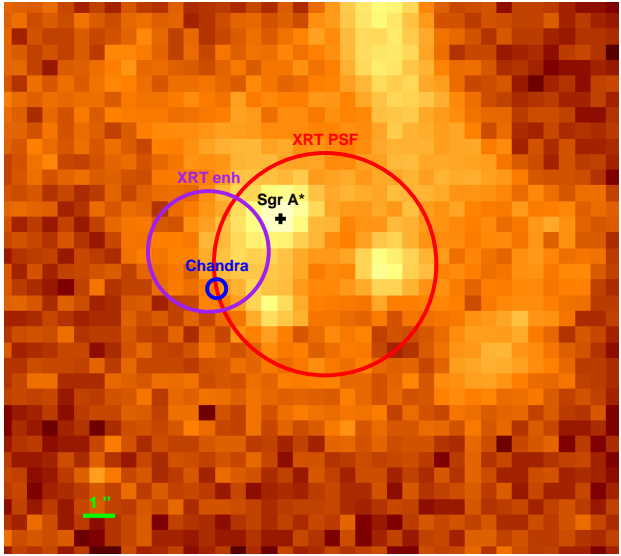}}
\caption{Localization error circles of \sgr. Shown here are the XRT
  PSF-fitted position (XRT PSF), the XRT position with astrometry
  correction (XRT enh), the \chandra\ position \citep{ReaATEL5032} and the
  radio position of \sgrastar. These error circles are over-plotted on a
  Chandra archival image of \sgrastar.}
\label{error_circles}
\end{figure}

The initial localizations of the XRT-detected point source reported by
\cite{BarthelmyGCN14443} and \cite{KenneaATEL5009} were consistent with the
position of both \sgrastar\ and the subsequent \chandra\ localization of \sgr. Using
field stars in the UV/Optical Telescope \citep{Roming05} field of
view, we improved the astrometry of the XRT position, using the method of
\cite{Goad07} and \cite{Evans09}.  We find a position of
$\alpha=17^\mathrm{h}45^\mathrm{m}40_{^.}^\mathrm{s}21$, $\delta=-29^\circ00^\prime29^{\prime\prime}_{^.}2$
with an uncertainty of $1^{\prime\prime}_{^.}9$ (radius, 90\% confidence). This error circle
rules out  this emission coming from \sgrastar, which lies $2^{\prime\prime}_{^.}5$
away, at $98\%$ confidence. The center of the \chandra\ error circle
\citep{ReaATEL5032} lies $1^{\prime\prime}_{^.}2$ from this position.  The relative
positions of the error circles are shown in Figure~\ref{error_circles}.

\subsection{Detection of the SGR Burst}

\begin{figure}
\resizebox{\hsize}{!}{\includegraphics[angle=270]{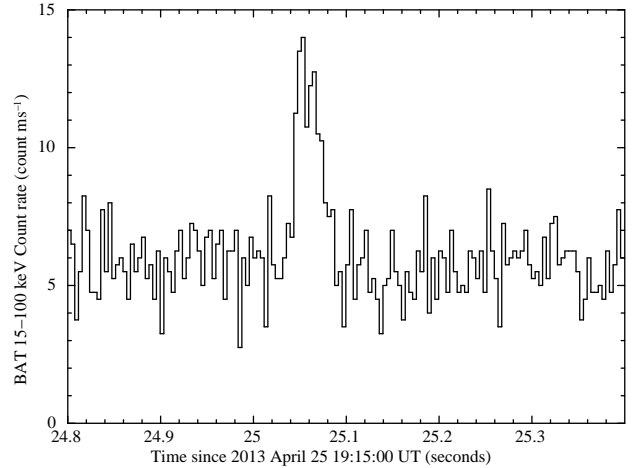}}
\caption{Light curve of the BAT detected burst from \sgr.}
\label{sgr_burst}
\end{figure}

The \swift\ BAT triggered on 2013 April 25 at 19:15:25 UT on a short hard
X-ray burst (Figure~\ref{sgr_burst}), detected at $10.4\sigma$
significance, from $\alpha=17^\mathrm{h}45^\mathrm{m}33^\mathrm{s}_{^.}3$,
$\delta=-28^\circ58^\prime55^{\prime\prime}_{^.}2$, with an uncertainty of
$2^{\prime}_{^.}1$ radius (90\% confidence, including systematic and statistical
errors). The XRT localization of \sgr\ lies marginally ($0'.05$) outside
this error circle. The burst consists of a single peak with duration of
$T_{90} = 0.028 \pm 0.009$ s.

The time-averaged spectrum of the burst is best fit by a single blackbody
model, with $kT_\mathrm{BB} = 9.2 \pm 0.8$\,keV ($\chi^2$ = 60.1 for 59
degrees of freedom); this corresponds to a blackbody emission region of
radius $1.5^{+0.4}_{-0.3}$\,km assuming a distance of 8\,kpc.  The burst
fluence in the BAT 15--150\,keV band was $7.8 \pm 1.8 \times 10^{-9}$ erg
cm$^{-2}$. A double-blackbody model, often favored for SGR bursts (e.g.,
\citealt{Lin12}), was not required to fit the BAT spectrum.

The characteristics of this burst are very similar to those of other SGR
bursts seen by BAT, e.g., those seen from SGR\,J1833$-$0832 \citep{Gogus10}
and Swift\,J1834.9$-$0846 \citep{Kargaltsev12}. As there exists no known SGR
within or near the BAT error circle, we conclude that this burst is from a
previously undiscovered SGR in the Galactic Center.

A scheduled observation of \sgrastar\ began at 19:14:10 UT, 75 s
before the BAT trigger. A search of the XRT light-curve from this
observation shows no evidence of an X-ray counterpart of the burst, with
only a single count in the 2.507\,s PC frame that covered the burst.  To
determine if the non-detection of the burst in XRT is consistent with the
BAT detection, we calculated the predicted number of X-ray counts that
would be seen in the 0.3--10\,keV XRT passband.  Using a model of an
absorbed blackbody ({\tt TBabs*bbodyrad}) with $N_{\rm H}$ set to the
average value (see Section~\ref{spec_av}), and the BAT fluence and
$kT_\mathrm{BB}$ values, we predicted $<1$\,counts from the burst,
consistent with its non-detection by XRT.

At the time of writing only one burst from \sgr\  has been seen by BAT. 
However, given that \sgr\ turned on between
25 and 50 hr before this burst, it is possible that there were earlier
bursts, not seen by BAT, which precipitated this turn-on.  We examined the
\swift\ observing plan to determine the BAT temporal coverage between the
XRT observations on April 23 and 24. 
During this time \sgr\ was only inside the BAT $>50\%$ coded field of
view $\sim4\%$ of the time, and therefore earlier bursts cannot be ruled out.

We have also searched for untriggered events from \sgr\ in the \fermi/Gamma
Ray Burst Monitor (GBM; \citealt{Meegan09}) Time-Tagged Event data with 16
ms time resolution. We did not find any burst in the data taken
pre-outburst or post-outburst from the \sgr\ direction. However, the search
also did not reveal any detection at the time of the BAT burst, suggesting
that GBM may be insensitive to such weak bursts.

\vskip 160pt
\subsection{\swift/XRT Spectral Analysis\label{spec_av}}

To characterize the XRT spectrum of \sgr\ we extracted a region centered on
the best fitted position in XRT detector coordinates of the transient,
using a radius of $10''$. This follows the method of \cite{Degenaar13}, to
maximize the signal from \sgr\ and minimize the effect of the bright
complex diffuse emission near \sgrastar. The background region was taken
from an annulus with inner and outer radii of $20''$ and $60''$,
respectively. We fit absorption using the {\tt TBabs} model with the
abundances set to the those of \cite{Wilms00} and the cross-sections set to
the values of \cite{Verner96}.

We fit the time-averaged spectrum for the longest single exposure taken
post-burst (ObsID 00554491001, with an exposure time of 19564\,s). The XRT
spectrum is dominated by the effects of high absorption, with negligible
X-ray emission below $\sim2$ keV.  The spectrum is well fit with either an
absorbed blackbody model or an absorbed power-law model. However, the
best-fit photon index for the power-law model ($\Gamma = 3.5\pm0.3$) is
very soft, suggesting that the spectrum is likely thermal in nature.

The parameters of the absorbed blackbody model are $N_\mathrm{H} =
13.7^{+1.3}_{-1.2} \times 10^{22}$~cm$^{-2}$ and $kT_\mathrm{BB} = 1.06 \pm
0.06$~keV ($\chi^2$ = 136.42 for 136 dof). The observed absorption is
consistent within errors with the quiescent spectrum of \sgrastar\ reported
by \cite{Nowak12}.  The addition of extra continuum components, e.g., the
blackbody plus power-law model often used to parameterize the spectra of
SGRs \citep{Kaspi10}, did not statistically improve the fit. However,
\nustar\ results show that the spectrum above 10\,keV does require a hard
power-law component to fit the data \citep{Mori13}.

The average observed flux in the 0.3--10\,keV band is $2.15^{+0.09}_{-0.08}
\times 10^{-11}$ erg s$^{-1}$ cm$^{-1}$ ($4.77^ {+0.40}_{-0.34} \times
10^{-11}$\,erg\,s$^{-1}$ cm$^{-2}$, corrected for absorption).  Assuming a
distance of 8\,kpc, this gives a luminosity of $3.6 \pm 0.3\times
10^{35}$\,erg\,s$^{-1}$ (0.3--10\,keV). The corresponding blackbody emission radius is equal
to $1.44^{+0.20}_{-0.16}$~km, with the caveat
that unfitted hard continuum components may be contributing to the XRT
flux.  We note, however, that this radius is consistent within errors to the
value derived from the BAT burst spectral fit.

\subsection{Investigation of Spectral and Flux Evolution}

To determine if there is any spectral or flux evolution detectable in the
\swift\ observations, we performed time resolved spectroscopy of the XRT
data in Table~\ref{observations}.  To maximize sensitivity to any changes
in the blackbody temperature and emission radius, we fixed the absorption
to the value reported in Section \ref{spec_av} and utilized Cash statistics
\citep{Cash79}, which generally provide more accurate fit parameters for
low counts spectra.

Because \swift\ is in a low Earth orbit, observations longer than $\sim
1.8$~ks are broken into multiple ``snapshots'', with start times separated
roughly by the \swift\ orbital period (96 minutes).  We extracted XRT
spectra for all snapshots longer than 100s.  To maximize the quality of the
data, we grouped adjacent snapshots within a single observation to achieve
a minimum exposure time of 2\,ks whenever possible.

We performed a similar analysis of the pre-burst data, with
$kT_\mathrm{BB}$ fixed to the value given in Section~\ref{spec_av}, and
calculated the 90\% confidence upper limit on the flux.

\begin{figure}
\resizebox{\hsize}{!}{\includegraphics[angle=270]{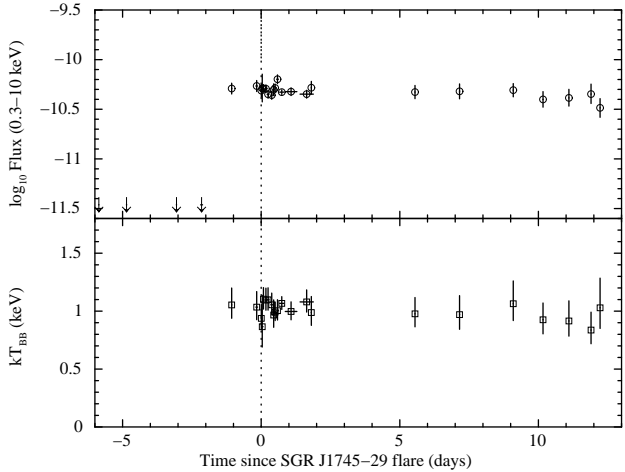}}
\caption{Upper panel: \swift/XRT flux evolution of \sgr, in the 0.3-10\,keV band, based on a single blackbody model fit 
corrected for absorption; upper limits (shown by arrows) based on the average spectral
  parameters are calculated for times before the outburst began in
  XRT. Lower panel: temperature ($kT$) evolution of time resolved spectra of the
  \swift/XRT data for a single blackbody model.}
\label{flux_and_temp}
\end{figure}

Absorption-corrected flux values (including upper limits) and $kT_\mathrm{BB}$ are
plotted in Figure~\ref{flux_and_temp}.  We find that all spectral
parameters are constant within errors, with no evidence of spectral
evolution, and an average $kT_{BB} = 1.02\pm0.04$\,keV. For observations
taken between 2013 April 25 and 2013 May 13 the flux is statistically
consistent with being constant, averaging $4.9 \pm 0.2 \times
10^{-11}$\,erg\,s$^{-1}$\,cm$^{-2}$. 
Long term monitoring will be necessary to constrain any decline in the
source flux.

\section{Discussion}

\swift\ has observed the sudden turn-on of a new transient source near
\sgrastar. This, combined with the BAT detection of a short hard X-ray
burst from a position consistent with the new transient, suggests this
transient is a new SGR in the Galactic Center, \sgr.  

\begin{figure}
\resizebox{\hsize}{!}{\includegraphics[angle=0]{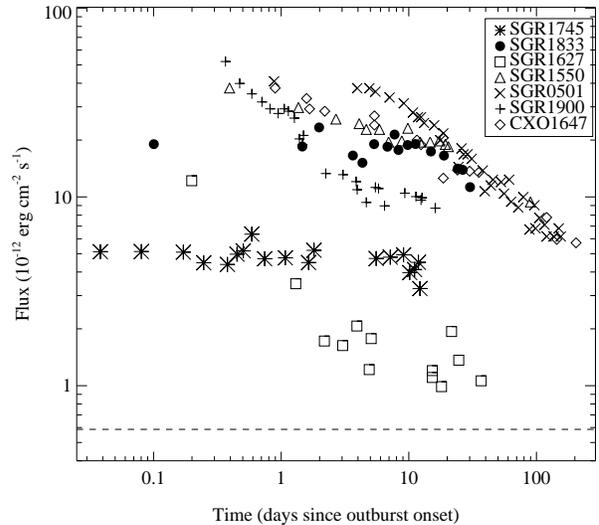}}
\caption{Absorption-corrected blackbody model flux of \sgr\ (stars)
  compared with other SGRs. Data are reproduced from \cite{Rea09} and
  \cite{Gogus10}.}
\label{flux_sgr_comp}
\end{figure}

The soft X-ray spectrum of \sgr, although hotter than the typical magnetar $kT_\mathrm{BB}
\sim0.5$\,keV value \citep{Woods06}, is consistent with the temperature
seen in some SGRs, for example Swift\,J1834.9$-$0846
\citep{Kargaltsev12}. Although we detect no evidence of fading, not all SGR
light-curves show detectable fading within weeks after the initial outburst
(e.g. SGR\,J1833$-$0832, \citealt{Gogus10}; see
Figure~\ref{flux_sgr_comp}).  Finally, \nustar, \chandra, \swift, and
several radio telescopes have measured a pulsar period of $\sim3.76$\,s,
consistent with the range of periods seen from SGRs, and reported
measurements of $\dot{P}$ \citep{GotthelfATEL5046,Mori13} suggest a
magnetic field of a few$ \times10^{14}$\,G, in the expected range for
magnetars \citep{Woods06}.

The $3''$ separation puts \sgr\ at a projected distance of
$\sim0.1$\,pc from \sgrastar.  Given the apparently similar absorption
column, we argue that \sgr\ is likely located close to \sgrastar.  At the
projected distance, the effects of \sgrastar\ on \sgr\ will be small: we
estimate that the gravitational acceleration due to \sgrastar\ will
contribute no more than $\pm 2 \times 10^{-13}$ to the apparent $\dot{P}$,
assuming nominal values of $4\times10^{6}\,M_\odot$ and 8\,kpc distance for
the central black hole.  We note that even with the constraints from
the absorption column, the true distance of \sgr\ from \sgrastar\ remains highly
uncertain.

As $\sim50\%$ of known magnetars lie within 30\,deg of the Galactic
Center, discovering a new SGR in this region area was not unexpected, but
the close proximity of this source to Sgr A*, combined with the temporal
coincidence with the anticipated encounter of G2 with Sgr A*, made this
event intriguing.  However, it seems unlikely that the turn-on of \sgr\ is
related to any interaction with G2, as G2 is currently within $1''$ of
\sgrastar\ \citep{Gillessen13}, whereas \sgr\ is $3''$ away. We conclude
that the onset of emission from \sgr\ at this time is coincidental.

The Galactic Center is very well studied in X-rays, allowing us to place
limits on burst and soft X-ray outburst emission from \sgr\ in the recent
past.  We estimate that \swift/BAT spends approximately 3.5\,Ms year$^{-1}$
covering the \sgrastar\ region, meaning that any repeated flaring activity
in the past eight years would likely have been seen.  \swift\ monitoring
observations of the Galactic Center region with XRT have been on-going
since 2006 February 24, so we can rule out any similar outburst with high
confidence for the past $\sim7$ yr.

\chandra\ has performed regular observations of this region starting 1999
September 21, and did not detect \sgr\ previously \citep{Muno09}. 

The excess diffuse emission, that is likely produced by colliding winds of
IRS 16SW and other nearby windy stars, makes it hard to estimate an upper
limit for the quiescent source state. However, \cite{Mori13} conservatively
estimate $\sim10^{32}$\,erg\,s$^{-1}$ (2-10 keV), based on the quiescent
limit on CXOGC\,J174540.0$-$290031 from \cite{Muno05}. We note that
SGR\,J1833$-$0832 was also observed pre-outburst by \chandra\ and was not
detected, with an upper limit of $3.4 \times
10^{-13}$\,erg\,cm$^{-2}$\,s$^{-1}$ \citep{Gogus10}, which is equivalent to
a luminosity of $4 \times 10^{32}$\,erg\,s$^{-1}$ for an assumed distance
of 5.7\,kpc, close to the \chandra\ limit on \sgr.

Since 2008 August, five new magnetar candidates have been discovered by
\swift\ and \fermi/GBM. Four of these have intriguing differences from the
previous members of the magnetar family: they are all transient sources
discovered by emitting typical magnetar short bursts, which became burst
inactive after exhibiting one or two relatively dim events, and their
persistent X-ray spectra are different than the rest of the magnetar
sources; they are typically well described by a single blackbody function
with a temperature around 1 keV ($0.3-10$ keV). \sgr\ shares these common
properties. In Figure~\ref{flux_sgr_comp} we present the unabsorbed flux trend of the
persistent X-ray emission from \sgr\ following the outburst onset, along
with that of a set of transient and persistent magnetars. It is striking to
note that the X-ray flux of both \sgr\ and SGR\,J1833$-$0832 remained fairly
constant in the first 10--20 days into the outburst, while that of other
transient magnetars (such as, SGR\,J1627$-$41 or SGR\,J1550$-$5418)
declined steadily following the outburst onset. We, therefore, suggest that
SGRs with low bursting rates possess slightly different characteristics
than the bulk of the population. We know from the spin and spin-down rates
of these sources that their dipole (or more local multi-pole magnetic
field) is in the magnetar regime. It is, however, possible that these
sources cannot efficiently radiate away the energy released by events
leading to bursts, therefore, cannot appear as prolific bursters. Instead,
the energy released in a burst event could be trapped within the system,
which could then result in crustal heating near the poles. It is, then
plausible that further energy release from the neutron star, possibly as
bursts, is continuously trapped, resulting in the constant persistent X-ray
flux seen in \sgr\ and other SGRs with apparent low bursting rates. In this
scenario, we would expect the \sgr\ flux to decline when the active episode
ends, typically after 1-2 weeks.

\acknowledgments

This work is supported by NASA grant NAS5-00135. This work made use of data
supplied by the UK Swift Science Data Centre at the University of
Leicester. We acknowledge the use of public data from the \swift\ data
archive. This research has made use of the XRT Data Analysis Software
(XRTDAS) developed under the responsibility of the ASI Science Data Center
(ASDC), Italy.

{\it Facility:} \facility{Swift}

\end{document}